\documentclass[runningheads]{llncs}
\usepackage[T1]{fontenc}
\usepackage{float}
\usepackage{tabularx}
\usepackage{amsmath}
\usepackage{amsfonts}
\usepackage{adjustbox}
\usepackage{multirow}
\usepackage{color}
\usepackage{longtable}
\usepackage{pdflscape}
\usepackage{rotating}
\usepackage{xcolor}

\usepackage{xcolor}
\usepackage[colorlinks = true,
            linkcolor = blue,
            urlcolor  = blue,
            citecolor = blue,
            anchorcolor = blue]{hyperref}
\usepackage[margin=1in,footskip=0.2in]{geometry}
\begin{document} 
\title{Modeling of Real and Imaginary Phase Shifts for
$\alpha-\alpha$ Scattering using Malfliet-Tjon Potential}

\author{Shikha Awasthi\inst{1}\orcidID{0000-0003-2939-638X} \and
Ishwar Kant\inst{1}\orcidID{1111-2222-3333-4444} \and
Anil Khachi\inst{2}\orcidID{0000-0002-6975-7357} \and
O.S.K.S. Sastri\inst{1}\orcidID{0000-0003-1405-5283}}
\authorrunning{Shikha Awasthi et al.}
%

\institute{Department of Physics and Astronomical Sciences, Central University of Himachal Pradesh,\\ Dharamshala-176215, HP, Bharat (India)\\
\and
Department of Physics, Chandigarh Group of Colleges, Jhanjeri,  Mohali-140307, Punjab, Bharat (India)}

\maketitle
\begin{abstract}
The real and imaginary scattering phase shifts (SPS) and potentials for $\ell=0,2,4$ partial waves have been obtained by developing a novel algorithm$^{\ref{Fig1}}$ to derive inverse potentials using a phenomenological approach. The phase equation, which is a Riccati-type non-linear differential equation, is coupled with the Variational Monte Carlo method. Comparisons between the resulting SPS for various $\ell$ channels and experimental data are made using mean absolute percentage error (MAPE) as a cost function. Model parameters are fine-tuned through an appropriate optimization technique to minimize MAPE. The results for $\ell=0^+$, $2^+$, and $4^+$ partial waves are generated to align with experimental SPS with mean absolute error (MAE) calculated with respect to experimental data is 3.19, 8.74, 13.06 respectively corresponding to real part and 0.76, 0.76, 0.59 corresponding to imaginary parts of scattering phase shifts.


\end{abstract}
\section{Introduction}\label{sec1}
\subsection{Background}
For many years, the research on of $\alpha-\alpha$ scattering has been a crucial field in nuclear physics, providing crucial information about the composition and dynamics of atomic nuclei. The development of scattering theory, especially the exploration of inverse problems, began in the 1950s. During this time, the concept of inverse problems emerged in several scientific domains, such as physics (i.e., electrodynamics, quantum scattering theory, and acoustics), geophysics (such as seismic and electromagnetic investigations), and astronomy. \\
In physics, especially in quantum scattering theory, inverse problems involve figuring out the potential or interaction that causes a specific scattering pattern \cite{Mackintosh}. This is different from direct problems, where the potential is known and the scattering pattern is determined. In $\alpha-\alpha$ scattering, the inverse problem aims to discover the nuclear potential that controls how two $\alpha$-particles interact.\\
With powerful computers, analyzing inverse problems has become easier, allowing detailed study of complex scattering processes. Modern computational techniques enable researchers to adjust potential models iteratively to match experimental data, which gives a better understanding of nuclear interactions. To properly describe scattering phenomena, the model parameters of interaction potential must be optimized.\\
Alpha particles, also known as helium nuclei, is made up of two protons and two neutrons. Their simple and stable nature makes them perfect for scattering experiments. Experiments on $\alpha$-particles illustrate the forces within the nucleus and also serve as a benchmark for theoretical models. Understanding $\alpha-\alpha$ scattering is crucial not only for nuclear physics but also for astrophysics, especially in stellar nucleosynthesis, where $\alpha$- particles are important in forming heavier elements \cite{1,2,3,4}. 
\subsection{Our previous results}\label{subsec1}
\subsubsection{SPS by Considering Morse potential}
Scattering phase shifts for the $\alpha-\alpha$ system within the elastic region were obtained in this work using the Morse potential with the Coulomb term included as an $erf()$ function. The SPS data in this work were taken from various experimental studies \cite{19,20,21}. Afzal et al. \cite{22} compiled experimental data for alpha-alpha scattering, which is commonly used in literature. Afzal $et al.$ limited the range of laboratory energies for SPS calculations to the elastic region of 0-23 MeV, or equivalently, 0-11.5 MeV center of mass (CM) energies \cite{22}. This paper aimed to include more accurate and precise experimental data from Chien and Brown, which covers the region from 18 to 29.50 MeV laboratory energies \cite{23}. The SPS from the optimized potentials for $\ell = 0^+$, $2^+$, and $4^+$ partial waves match with the experimental SPS for experimental energies up to 25.5 MeV, with MAPE values of 1.17, 0.69, and 1.77, respectively.
\subsubsection{SPS by Considering Double Gaussian potential}
In this paper, scattering phase shifts for the $\alpha-\alpha$ system were obtained using a two-term Gaussian potential along with the Coulomb term represented by an $erf()$ function, within the elastic region ($E_\ell = 0-23$ MeV). The PFM method was employed for all even partial waves: $\ell = 0, 2, 4$, as well as $\ell = 6, 8, 10$. Although their impact on the total cross-section is small, these partial waves were included because they cannot be overlooked. Additionally, calculations were expanded to higher $\ell$ channels such as $\ell = 6, 8,$ and $10$, which were not considered in Ali $et al.$'s work. The results for $\ell = 0^+$, $2^+$, and $4^+$ partial waves were obtained to match experimental SPS, yielding MAPE values of $2.9$, $4.6$, and $6.2$, respectively, for data up to $23$ MeV. For higher states $\ell = 6^+$, $8^+$, and $10^+$, the MAPE values were $3.2$, $4.5$, and $5.9$, respectively, for data from $53-120$ MeV. The PFM method was used in conjunction with a technique for optimizing model parameters to obtain the results.
\subsection{Complex potential approach}\label{subsec2}
For scattering energies above the breakup threshold, the imaginary part of scattering phase shifts (SPS) grows very rapidly and hence contributes significantly to total scattering cross sections \cite{Huber}. To calculate the imaginary parts of the scattering phase shifts, we followed the approach of Darriulat $et al.$ \cite{Darriulat} and A.K. Jana \cite{Jana}, who used the concept of complex potential to compute both the real and imaginary parts of scattering phase shifts. They obtained imaginary SPS for $\alpha-\alpha$ scattering by considering complex Woods-Saxon potential as the interaction potential.\\
Darriulat \cite{Darriulat} studied differential cross section of elastic scattering between $\alpha$ particles and $He^4$ across lab energies ranging from 53 to 120 MeV, analyzing it in terms of complex phase shifts. Jana \cite{Jana} derived phase equations for S, D, and G waves (corresponding to $\ell=0,2,4$) in $\alpha-\alpha$ scattering using the Green’s function approach. They conducted a phase shift analysis for both real and imaginary scattering phase shifts (SPS) using the phase function method (PFM) \cite{Calogero}. G.R. Satchler $et.al.$ \cite{Satchler} investigated nucleon-$\alpha$ elastic scattering below 20 MeV using a real optical model potential in Woods-Saxon form and employed a complex central potential for energies exceeding 30 MeV. In all these studies, the Coulomb potential was modelled as that generated by a uniformly charged sphere with a radius $r_C$. S. Ali $et.al.$ \cite{Ali} applied a phenomenological potential approach to $\alpha-\alpha$ scattering, utilizing a modified Coulomb potential based on the $erf()$ function. 
\subsection{Outline of current work}\label{subsec3}
\begin{figure*}[htp]
\centering
{\includegraphics[scale=0.4]{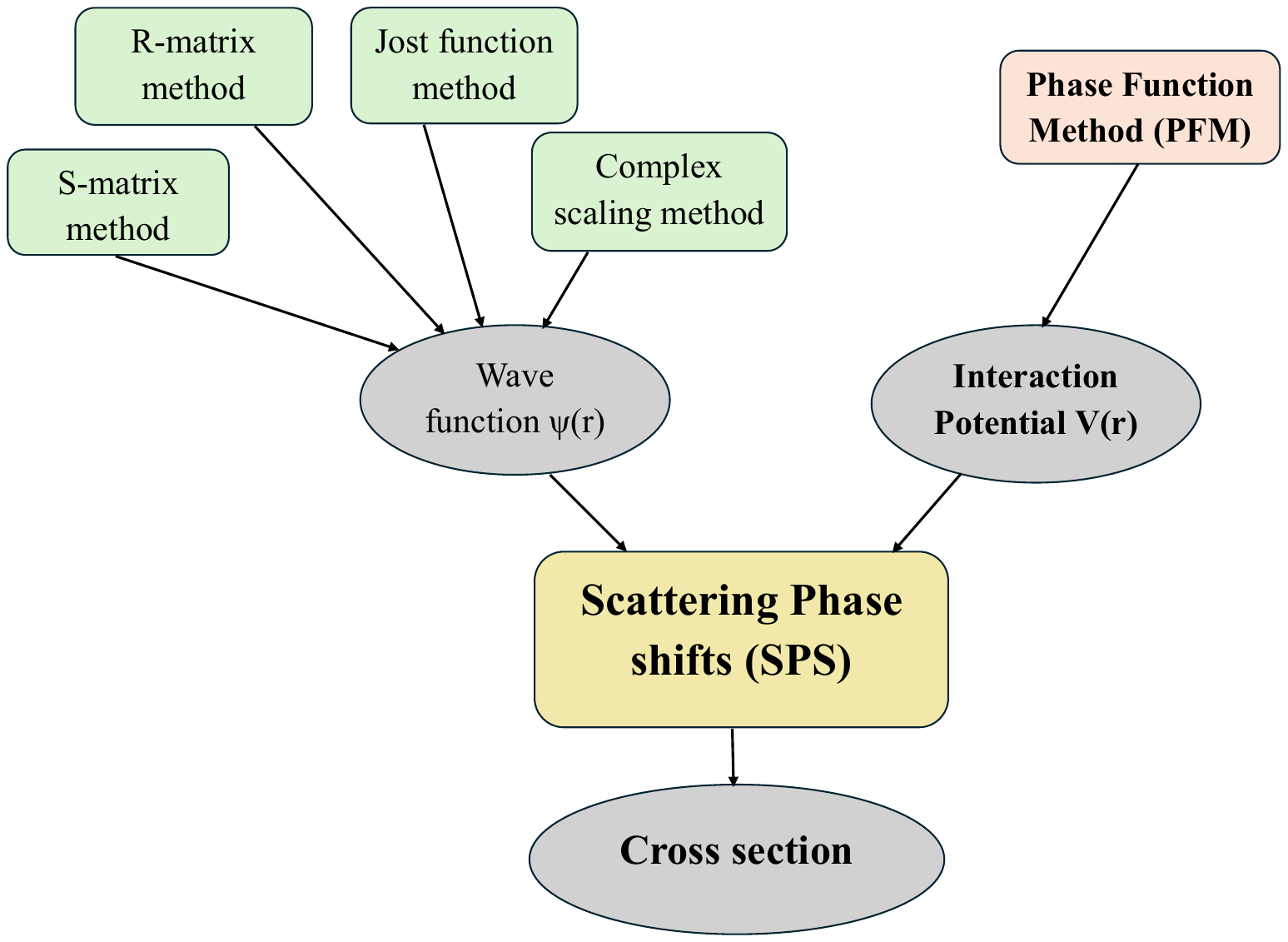}}
\caption{Different methods to calculate cross section Importance of phase function method (PFM).}
\label{fig4}
\end{figure*}

In this work, we have obtained scattering phase shifts by using the phase function method (PFM) \cite{Calogero} using inverse potential approach \cite{PRC}. The PFM is an effective method for calculating scattering phase shifts, and it works very well for local potentials that tend to zero very quickly as $r$ increases. In the PFM approach, the time-independent Schr$\ddot{o}$dinger equation (TISE) is transformed into a first-order non-linear Ricatti-type equation that directly deals with phase shifts for different $\ell$ values and different energies for a chosen interaction potential, without the need for the wave function. On the other hand, S-matrix method \cite{smatrix}, R-matrix method \cite{rmatrix} or Jost function method \cite{jost} etc. rely on wave function to obtain SPS for various partial waves. This can be graphically seen in figure \ref{fig4}.\\
Since in PFM, only the interaction potential is required to obtain scattering phase shifts and the final phase shift is calculated at the asymptotic region where the potential does not produce any phase shift \cite{Calogero}. So, to make the concept of phase shifts well-defined, it becomes important that the potential is well-behaved and vanishes smoothly at a particular distance r$>$R. \\
In our previous work mentioned earlier in section \textbf{1.2}, we obtained scattering phase shifts and corresponding interaction potentials for $\alpha-\alpha$ scattering by considering Morse potential and double Gaussian potential for explaining the nuclear part. For the Coulomb part, we considered $erf()$ function in both cases. But the $erf()$ function has limitations regarding the integration limits because $erf()$ function has long-range nature and is not well-defined, therefore, we have to consider a short-range potential which serves as a screened Coulomb potential \cite{Preston}.\\ 
The ambiguity of long-range Coulomb potential can be avoided by considering Hulth{\'e}n potential, as considered by Laha $et al.$ \cite{AtHulthen}, which behaves as a screened Coulomb potential at short distances and dies down exponentially at large distances. The Hulth{\'e}n potential is an important short-range potential in physics. Laha $et al.$ have successfully applied nuclear Hulth{\'e}n potential to study $\alpha-nucleon$ \cite{Bhoi}, $\alpha-\alpha$ and $\alpha-He^3$ \cite{Laha3} elastic scattering.\\
Typically, the long-range character of Coulomb potential has been overcome by considering the screened Coulomb potential of Hulth{\'e}n form \cite{Xin}. This is acceptable considering that the target is typically polarised due to scattering and is hence surrounded by residual particles. Laha $et.al.$ \cite{Laha3} have utilised atomic Hulth{\'e}n form as an ansatz for screened Coulomb potential to study $\alpha-nucleon$, $\alpha-\alpha$ and $\alpha-He^3$ elastic scattering.\\
In place of complex Woods-Saxon potential, we consider Malfliet-Tjon potential (MT) \cite{MT} as an interaction function and obtained SPS for $\ell=0,2,4$ states of $\alpha-\alpha$ scattering. The phase shifts for real and imaginary parts have been obtained for laboratory energies ranging from 1 to 120 $MeV$ utilising phase function method (PFM) \cite{Calogero} using phenomenological potential approach \cite{Selg}.\\
In the next section, we will provide a literature review of $\alpha-\alpha$ scattering. In Section \textbf{3} methodology to obtain scattering phase shifts is given, results and discussion are presented in Section \textbf{4} and finally, the conclusion is drawn in Section \textbf{5}.
\section{$\alpha-\alpha$ scattering}
\subsection{Review of the Earlier $\alpha-\alpha$ Scattering Results}
Rutherford and Chadwick were the first to experimentally study $\alpha-\alpha$ scattering in the year 1927 and since then a large amount of experimental data has been available, given by (i) Afzal \textit{et al.} \cite{Afzal} (ii) S. Chien and Ronald E brown \cite{Chien} (iii) Igo \cite{Igo} (iv) Darriulat, Igo, Pug, Holm \cite{Darriulat}(v) Nilson \cite{Nilson} and others. alpha-alpha problem has been extensively studied both experimentally and theoretically, with alpha particle having some sole properties like (i) zero spin and isospin, (ii) tight binding energy of $28.3$ MeV, having property to form cluster-like states for lighter nuclei ($^{6,7}$Li, $^9$Be, $^{12}$C and $^{16}$O are $\alpha$-structured) with alpha particle being the core nuclei in the cluster (iii) small root mean square radius of $1.44$ fm.\\
In the 1940's for $\alpha-\alpha$ scattering experiments, only naturally occurring $\alpha$-sources like polonium, thorium and radium were used, which did not result in very accurate results. Later on, with the advancement of technology, accelerators were used in scattering processes and highly accurate phase shifts were observed. The importance of $\alpha-\alpha$ scattering is that the study provides information regarding the force field in the vicinity of He-nuclei and also provides information regarding energy levels of $^8$Be nucleus.\\ 
Haéfner was one of the first to study $^8$Be properties in 1951 by using phenomenological potential \cite{Hafner}. Later on, Nilson, Briggs, Jentschke and others used Haéfner potential for $\ell = $2 state from the ground state of $96$~KeV. Later on, Nilson \textit{et al.} \cite{Nilson} extended Hafner model \cite{Hafner} to include $\ell$= 0, 2 and 4 and found that best agreement of phase shift with experiments requires small value of $R = 3.49$ fm. Also, Nilson found that testing the efficiency of the potential required data above $22.9$ MeV, which was unfortunately not available at that time. Later in the year 1958 Spuy and Pienaar \cite{Van} made phenomenological analysis up to $6$ MeV, where they concluded that for $E<6$ MeV one needed velocity-dependent interaction for fitting S and D waves. Later on, Wittern \cite{Wittern} in the year 1959 derived the same semi-phenomenological potential and reached the same conclusion as given by Spuy and Pienaar. Later, from 1960 to 1965, more phenomenological study was done by Igo who made an optical model analysis for $\alpha-\alpha$ in range $E = 23.1-47.1$ MeV and Darriulat, Igo and Pugh \cite{Igo} who used energy independent but strongly $\ell$ dependent complex Wood Saxon potential for range $E = 53-120$~MeV where they failed to fit the phase shifts using one single common potential for all the partial waves. Thus, it has been concluded, that a single potential common to all $\ell$ do not exist phenomenologically. In short, $\alpha-\alpha$ potentials are found to be strongly $\ell$ dependent.
\subsection{More Recent $\alpha-\alpha$ Scattering Results}
In nuclear physics, the understanding of nuclear forces is very important to explore the stability and behaviour of nuclei. Our understanding of radioactivity has been significantly increased by Rutherford and his team's pioneering research \cite{Preston}. They found that the $\alpha$ particle is the helium-4 nucleus, with unique properties like $0^+$ spin and parity, Bose-Einstein statistics, a radius of 1.44 fm, and a binding energy of 28 MeV. These discoveries are fundamental to modern nuclear physics and the role of $\alpha$-particles. Early theoretical studies aimed to find out if the interaction between two $\alpha$-particles includes short-range repulsion and attraction over 3–4 fm, which is the typical distance between them in a nucleus \cite{Afzal}. This repulsion prevents the $\alpha$-particles from getting too close and overlapping. The repulsive interaction between $\alpha$-particles in $^8Be$ depends on velocity and can not be represented graphically. Yet, the extent of overlapping of $\alpha$-particles in $^8Be$ can be understood by plotting the effective phenomenological $\alpha-\alpha$ potentials.\\
Serdar Elhatisari $et. al.$ \cite{Serdar} in 2015 used lattice effective field theory to describe the low-energy interactions of protons and neutrons, and applied a technique called the ‘adiabatic projection method’ to reduce the eight-body system to a two-cluster system and to calculate \textit{ab initio} effective Hamiltonian for the two clusters. In 2022, Serdar Elhatisari $et. al.$ investigated the phase shifts of low-energy $\alpha-\alpha$ scattering under variations of the fundamental parameters of the Standard Model. E. Ruiz Arriola \cite{Ruiz} in 2008, discussed $\alpha-\alpha$ scattering in terms of a chiral two pion exchange potential (TPE) and concluded that when $^8Be$ is treated as a resonance state, a model-independent correlation between the Q‐factor and lifetime for the decay into two alpha particles arises.\\
Ngo Hai Tan \cite{Ngo} $et al.$ in 2014, did the folding model analysis of the elastic $\alpha-\alpha$ scattering at the incident energies below the reaction threshold of 34.7 MeV (in the lab system), using realistic density-dependent M3Y interaction. In 2021, Y. Hirabayashi \cite{Hirabayashi} studied the $\alpha-\alpha$ cluster structure at the highly excited energy in $^8Be$ in the coupled-channel calculations using the double-folding model and showed the existence of the $\alpha-\alpha$ cluster structure at the highly excited energy around $E_x=20$ MeV. B. Apagyi \cite{Apagyi} in 2022, applied the fixed-energy quantum inversion method to determine model independent $\alpha$-particle scattering potentials within the energy range between 0 and 50 MeV by considering only even experimentally allowed phase shifts. M. Odsuren $et.al.$ \cite{Odsuren} in 2023, used the complex scaling method (CSM) to study scattering phase shifts and resonance contributions in different two-body systems, examining its effectiveness with various potential models. S. Chakraborty $et.al.$ \cite{Chakraborty}, performed a simultaneous phenomenological R-matrix analysis using measured capture reaction cross sections, elastic excitation function, and phase shift data.
\section{Methodology}\label{sec2}
\subsection{Modeling the interaction}
Considering the $\alpha-\alpha$ system to be a simplified one-body system, the phenomenological interaction potential would consist of nuclear and Coulomb contributions. Since we are considering energies beyond 25 MeV, one has to determine both real and imaginary SPS separately, by choosing appropriate mathematical functions for the real and imaginary parts of a complex potential. Hence, we have considered a complex potential model as suggested by Darriulat \cite{Darriulat} based on Malfliet-Tjon (MT) potential, given by:
\begin{equation}
V_{N}(r) =  -V_{AR}\Big(\frac{e^{-\mu_{AR} r}}{r}\Big)+ V_{RR}\Big(\frac{e^{-\mu_{RR} r}}{r}\Big) - i \Bigg( -V_{AI}\Big(\frac{e^{-\mu_{AI} r}}{r}\Big)+ V_{RI}\Big(\frac{e^{-\mu_{RI} r}}{r}\Big)\Bigg)
\label{Opticalpot}
\end{equation}
Where, $\mu_R = 2\mu_A$ in units of $fm^{-1}$. Since, $\alpha-\alpha$ is a charged system therefore in addition to the nuclear interaction, one needs to account for Coulomb interaction due to protons. In this work, we have considered a screened potential as suggested by Laha $et.al.$ \cite{AtHulthen}, as Coulomb potential. This potential is a modified form of Yukawa potential and dies exponentially as Coulomb potential, given by
\begin{equation}
V_C(r) = V_0 \frac{e^{-r/a}}{1-e^{-r/a}} 
\label{vc}
\end{equation}
Here $V_0$ is the strength of the potential and parameter $a$ is chosen in such a way that $aV_0=2k \eta$. Here $\eta$ is a constant quantity known as the Sommerfeld parameter given by
\begin{equation}
\eta= \frac{\alpha}{\hbar v}
\end{equation}
Where, $\alpha=Z_1Z_2e^2$ and $v=\sqrt{\frac{2E}{\mu}}$ is the relative velocity of reacting particles at large separation.\\
Hence, $\eta$ after all the substitutions will finally be given by
\begin{equation}
\eta=\frac{Z_1Z_2e^2 \mu}{\hbar^2 k}
\end{equation}
Therefore, $aV_0$ after multiplying and dividing by $c^2$ will be equal to
\begin{equation}
aV_0=2k \eta = \frac{2 Z_1Z_2e^2 \mu c^2}{\hbar^2 c^2} = 0.2758~~~fm^{-1}
\end{equation}
Here, $\hbar c = 197.327$~MeV-fm, $e^2 = 1.44$~MeV-fm and $\mu c^2$ is the reduced mass of the system in the units MeV/$c^2$, hence, the value of $aV_0$ is a fixed value for any particular interaction. Adding screened Coulomb potential to our nuclear potential (MT potential) thus makes the total interaction potential well-defined \cite{AtHulthen} and can be put in the phase equation for obtaining scattering phase shifts.
\subsection{Phase Function Method:}
The Schr$\ddot{o}$dinger wave equation for a spinless particle with energy E and orbital angular momentum $\ell$ undergoing scattering with interaction potential V(r) is given by
\begin{equation}
\frac{\hbar^2}{2\mu}\bigg[\frac{d^2}{dr^2}+\bigg(k^2-\frac{\ell(\ell+1)}{r^2}\bigg)\bigg]u_\ell(k,r)=V(r)u_\ell(k,r)
\end{equation}
Where
\begin{equation}
k_{c.m}=\sqrt{\frac{2\mu E_{c.m}}{{\hbar^2}}}~fm^{-1}
\end{equation}
with $\frac{\hbar^2}{2\mu}=10.44217$ MeV $fm^2$. For $\alpha-\alpha$ system, center of mass energy $E_{c.m.}$ is related to laboratory energy by following the relation for non-relativistic kinematics 
\begin{equation}
E_{c.m}=\frac{M_\alpha}{M_\alpha+M_\alpha}E_{\ell ab}=0.5E_{\ell ab}
\end{equation}
The phase function method (also known as the Variable phase approach) is an important method for studying scattering in both local \cite{Brazillian} and non-local \cite{Jana} interactions. The method is based on well-known math principles, which show that a second-order equation like the Schr$\ddot{o}$dinger equation can be simplified to a first-order nonlinear differential equation (NDE) called the Riccati equation \cite{Morse}. This phase equation was independently developed by Calogero \cite{Calogero} and Babikov \cite{Babikov} and is given below.
\begin{equation}
\delta_{\ell}'(k,r)=-\frac{V(r)}{k(\hbar^2/2\mu)}\big[\cos(\delta_\ell(r))\hat{j}_{\ell}(kr)-\sin(\delta_\ell(r))\hat{\eta}_{\ell}(kr)\big]^2
\label{PFMeqn}
\end{equation}
This NDE is numerically integrated from the origin to the asymptotic region using suitable numerical techniques, thereby directly obtaining the values of the scattering phase shift for different values of projectile energy in a laboratory frame. The phase shift (solution of a non-linear differential equation) is the limiting value
$\delta=\lim_{r \to \infty} \delta_\ell(r)$. The central idea of VPA is to obtain the phase shift $\delta$ directly from physical quantities such as interaction potential V(r), instead of solving TISE for wave functions u(r), which in turn are used to determine $\delta_{\ell}(k,r)$. With initial condition $\delta(0)=0$.\\
The phase shift $\delta_{\ell}$ can be seen as a real function of $k$ and characterizes the strength of scattering of any partial wave, i.e. say $\ell^{th}$ partial wave of the potential V(r). In the above equation and are the Bessel functions. Since we are only focusing on obtaining scattering phase shifts for $\ell$= 0 partial wave, the Riccati-Bessel function \cite{Calogero} is given by $\hat{j_{0}}=\sin(kr)$ and similarly the Riccati-Neumann function is given by $\hat{\eta_{0}}=-\cos(kr)$, thus reducing eq. \ref{PFMeqn} to
\begin{align}
  \delta_0'(k,r) &= \begin{aligned}[t]
      &  -\frac{V(r)}{k(\hbar^2/2\mu)}\bigg[  \sin(kr+\delta_0)\bigg]^2 \\
       \end{aligned}
\intertext{PFM equation for D-wave takes the following form}       
  \delta_2'(k,r) &= \begin{aligned}[t]
      &  -\frac{V(r)}{k(\hbar^2/2\mu)}\bigg[  -\sin{\left(kr+ \delta_2 \right)}-\frac{3 \cos{\left(\delta_2 + kr \right)}}{kr} + \frac{3 \sin{\left(\delta_2 + kr \right)}}{k^{2} r^{2}}\bigg]^2 \\
      &  
       \end{aligned}
\intertext{PFM equation for G-wave takes the following form}
  \delta_4'(k,r) &= \begin{aligned}[t]
  \begin{cases}
&-\frac{V(r)}{k(\hbar^2/2\mu)}\bigg[\sin{\left(\delta_4 + kr \right)} + \frac{10 \cos{\left(\delta_4 + kr \right)}}{kr}-\frac{45 \sin{\left(\delta_4 + kr \right)}}{k^{2} r^{2}}- \frac{105 \cos{\left(\delta_4 + kr \right)}}{k^{3} r^{3}}\\
&+ \frac{105 \sin{\left(\delta_4 + kr \right)}}{k^{4} r^{4}}\bigg]^2
      \end{cases}
       \end{aligned}
\end{align}
These NDE’s equations (Eq. 10-21) are numerically integrated from the origin to the asymptotic region using the RK-4/5 method, thereby directly obtaining the values of the scattering phase shift for different values of projectile energy in the lab frame. The central idea of phase function method is to obtain the phase shift $\delta$ directly from physical quantities such as interaction potential V(r), instead of solving TISE for wave functions $u(r)$, which in turn are used to determine $\delta$. 
\subsection{Optimisation of model parameters using Variational Monte-Carlo (VMC) method:} 
The Variational Monte Carlo (VMC) method is a powerful technique that blends two important approaches: the randomness of Monte Carlo simulations and the careful optimization of variational methods. This combination helps to explore a system's configuration efficiently. In VMC, we use the randomness of Monte Carlo simulations to iteratively adjust the model parameters, getting closer to the desired configuration step by step. In our approach, we change potential model parameters based on experimental data. Below are the steps and block diagram \ref{Fig1} for implementing VMC on a computer:
\begin{figure*}[htp]
\centering
{\includegraphics[scale=0.56]{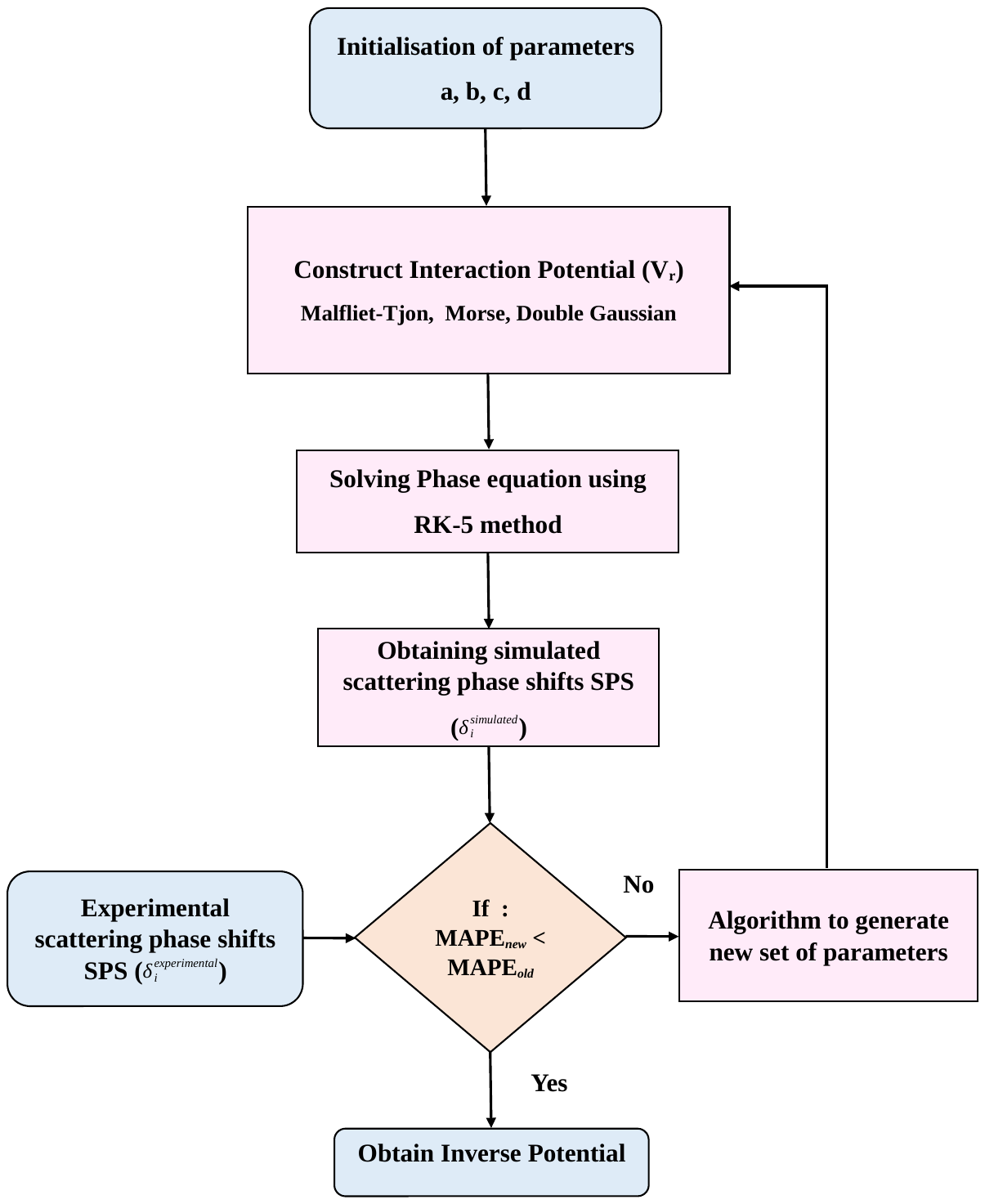}}
\caption{Block diagram to obtain inverse potentials for $\alpha-\alpha$ scattering.}
\label{Fig1}
\end{figure*}
\begin{enumerate}
\item \textbf{Initialization of model parameters:}\\
In this step, we start by choosing the initial values for the parameters of the interaction potential $V_r$ say, $a$, $b$, $c$ and $d$, based on theoretical or empirical data. 
\item \textbf{Solving PFM equation using RK method:}\\
In this step, we numerically solve the PFM equation (Eq. 37) using the Runge-Kutta method (specifically, the RK-5 method used here) to obtain the simulated scattering phase shifts (SPS), which we call $\delta_{old}$. We then calculate the mean absolute percentage error (MAPE) between the simulated SPS and the experimental data and record it as $MAPE_{old}$.
\item \textbf{Monte Carlo step:}\\
In this step, a random number '$r$' is generated within the interval [-I, I]. This random number is then added to one of the four parameters, such as $a_{new} = a + r$.
\item \textbf{Recalculating the Scattering Phase Shifts:}\\
The scattering phase shift is recalculated using a new set of perturbed parameters, namely $a_{new}$, $b$, $c$, and $d$. The mean absolute percentage error is then recalculated and saved as $MAPE_{new}$.
\item \textbf{Variational step:}\\
In this step, we verify whether the condition $MAPE_{new} < MAPE_{old}$ is met. If it is, the parameter $a$ is replaced with the updated value $a_{new}$; otherwise, the old value is preserved.
\item \textbf{Iterative steps:}\\
Keep repeating steps 3, 4, and 5 for all parameters to finish one iteration. Then, reduce the size of interval '$r$' after a certain number of iterations to see if MAPE decreases further. Repeat this process until MAPE stops changing, indicating convergence.
\end{enumerate}
\section{Results and Discussions}\label{sec4}
\begin{figure*}[htp]
\centering
{\includegraphics[scale=0.51]{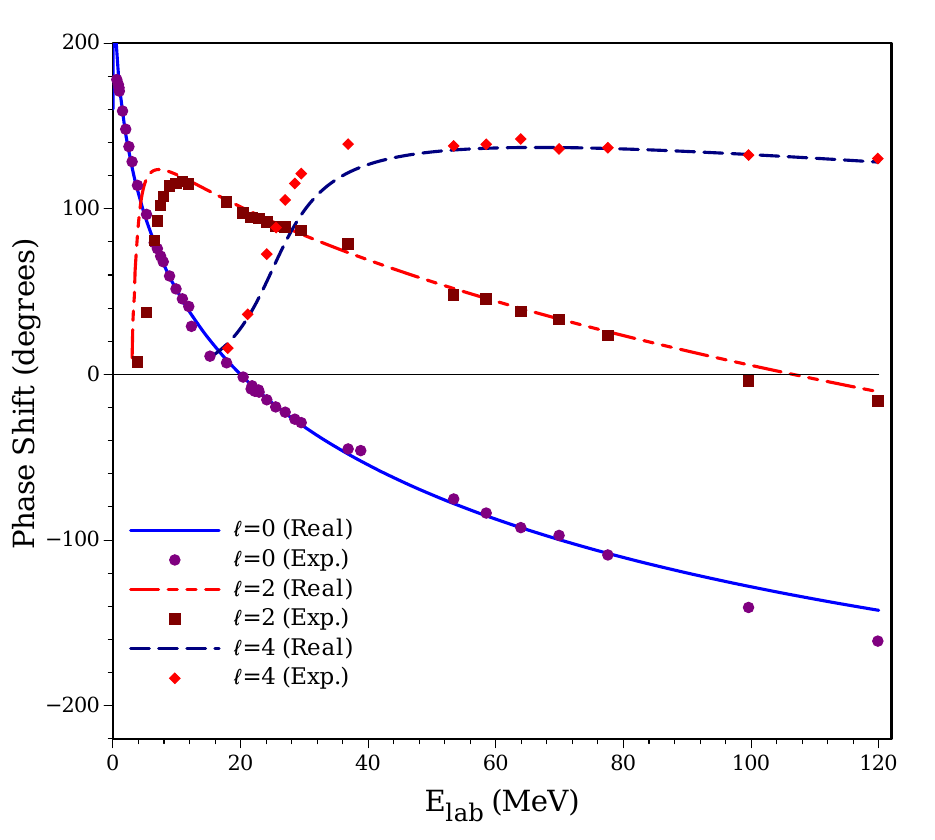}}
{\includegraphics[scale=0.5]{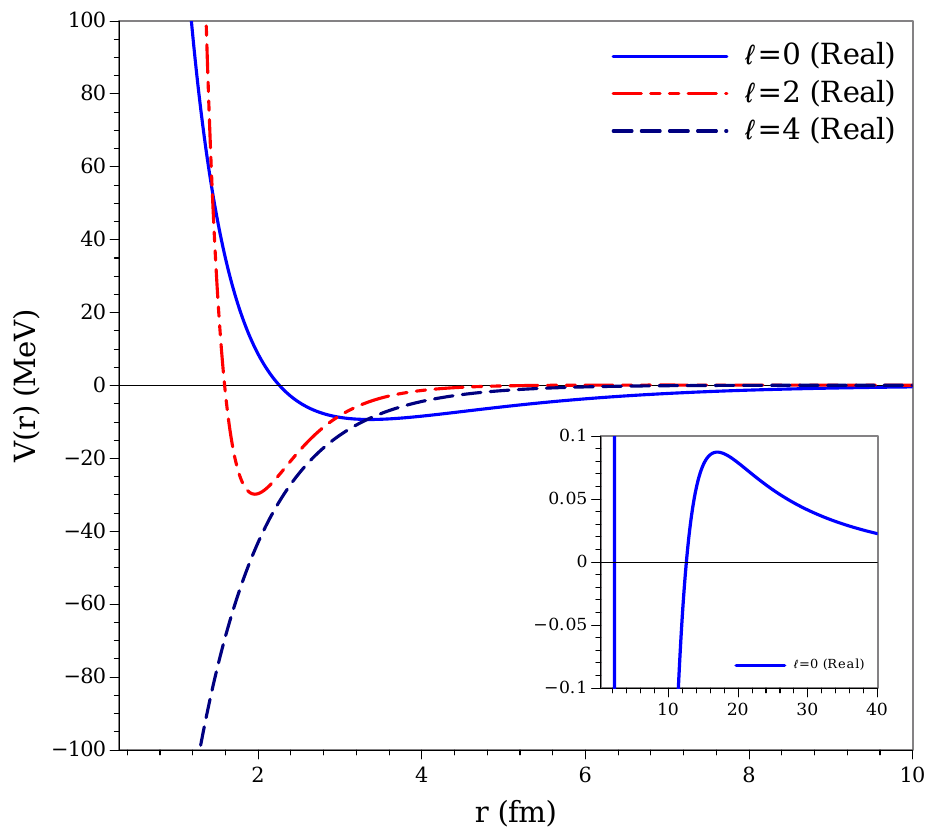}}
\caption{Real scattering phase shifts (left) and corresponding potential plots (right) for $\ell = 0,2,4$ channels of $\alpha-\alpha$ scattering. The inset figure shows Coulomb barrier height for $\ell=0$.}
\label{real}
\end{figure*}

\begin{figure*}[htp]
\centering
{\includegraphics[scale=0.5]{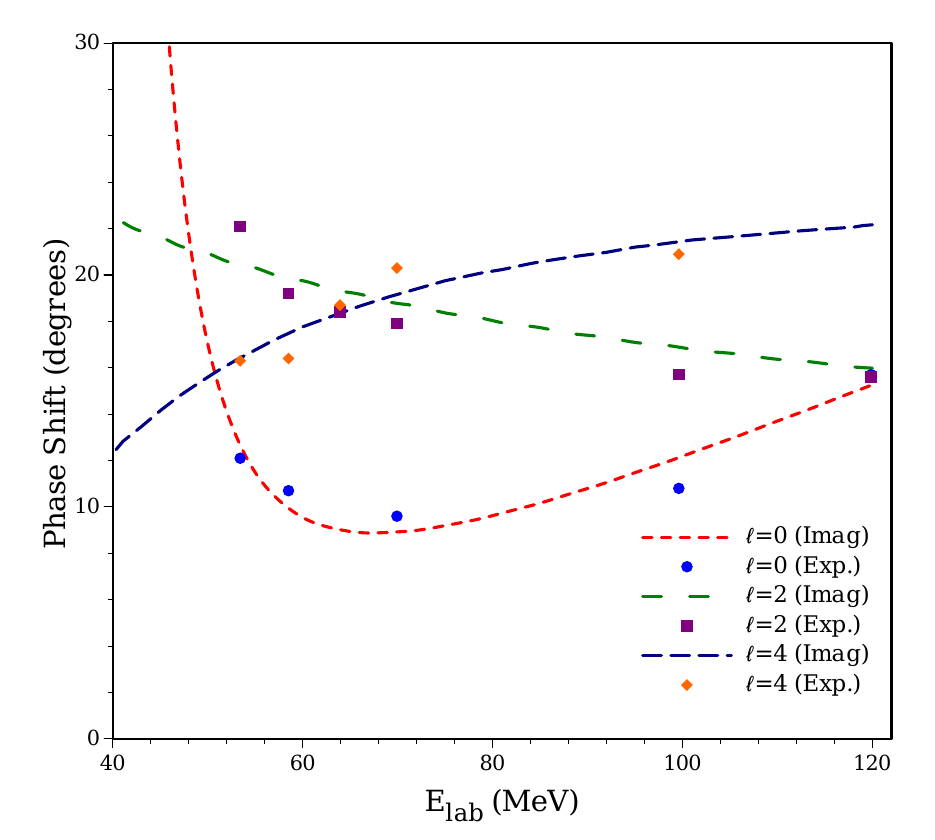}}
{\includegraphics[scale=0.5]{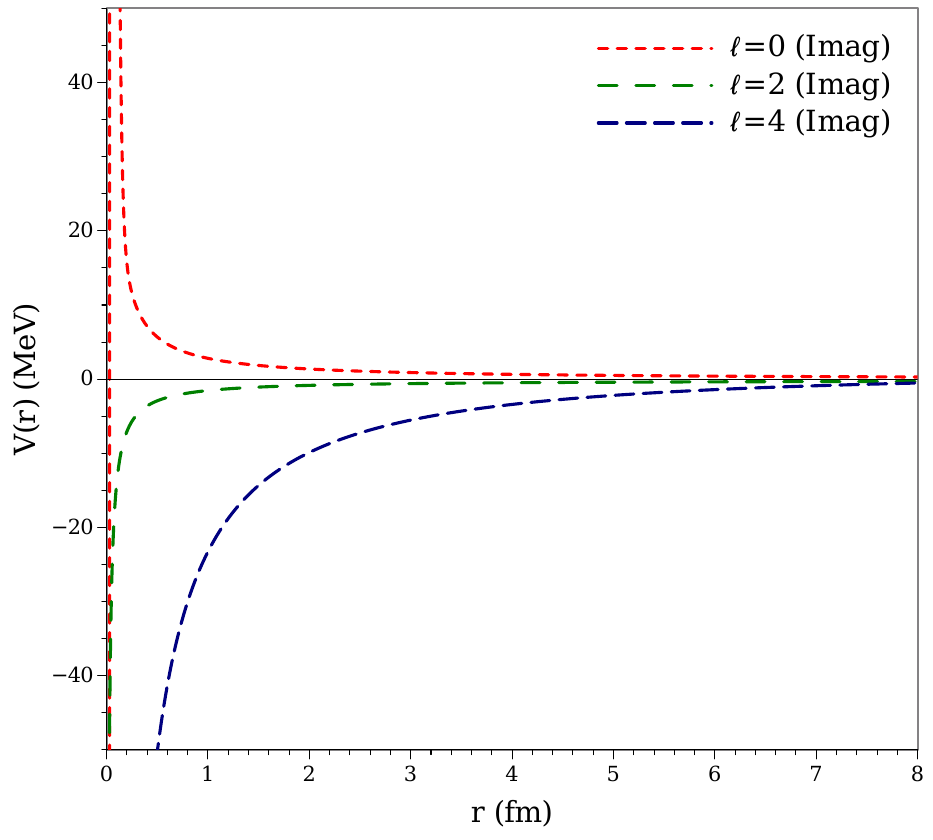}}
\caption{Imaginary scattering phase shifts (left) and corresponding potential plots (right) for $\ell = 0,2,4$ channels of $\alpha-\alpha$ scattering.}
\label{imag}
\end{figure*}

In this work, we have considered Malfliet-Tjon potential defined in equation Eq. \ref{Opticalpot}, to calculate real and imaginary parts of scattering phase shifts for $\alpha-\alpha$ scattering using phase function method. Hulth{\'e}n potential defined in equation Eq. \ref{vc} takes care of long range electromagnetic interaction between the charged particles. \\
In phase equation Eq. \ref{PFMeqn}, the function $\delta'_{\ell}(k,r)$ is referred to as the phase function. Its value at r = R gives the phase shift for the interaction potential V(r) at that point. The phase shift corresponding to any $\ell^{th}$ partial wave is determined by the phase function $\delta'_{\ell}(k,r)$ as r approaches infinity as $\delta_{\ell}(k)= \lim_{r\to\infty}\delta_{\ell}(k,r)$. We have optimised the model parameters of total interaction potential by following the procedure given in figure \ref{Fig1}. For all simulated phase shifts by minimising MAPE, we have calculated mean absolute error (MAE) corresponding to all channels with respect to experimental data.\\
The model parameters along with screened parameter values for real and imaginary parts of SPS are taken as $\ell = 0,2,4$ channels (S, D, G states) are given in Table \ref{parameters} below. The values of screened Coulomb potential for calculating both real and imaginary phase shifts corresponding to S, D, G states have been taken work done by our group previously \cite{Ayushi}. In the last column, calculated MAE for all the states is given.
\begin{table}[h!]
\centering
\caption{Model parameters: $V_{R}$ $\small{(MeV-fm)}$, $V_{A}$ $\small{(MeV-fm)}$, $\mu_{A}$ $\small{(fm^{-1})}$ for MT potential and mean absolute error (MAE) w.r.t expected data, for real and imaginary parts of $\ell = 0,2,4$ states for $\alpha-\alpha$ scattering.}
\begin{tabular}{p{2cm}|p{2.5cm}|p{2.5cm}p{2.5cm}p{2.5cm}p{2cm}|p{2cm}}
\hline \hline
~~SPS & ~~~~~State &~~~~$V_r$ &~~$V_a$ &~~$\mu_A$ & ~~~$a$ & ~~~~MAE\\
&&\small{(MeV-fm)}& \small{(MeV-fm)} & ($fm^{-1}$) & (fm) & \\
\hline
\multirow{2}{*}{~~Real} & ~~~~$\ell$ = 0  & ~801.305 & ~338.413 & 0~.386 & ~~~20 & ~~3.19\\
& ~~~~$\ell$ = 2 & ~28539.229 & ~2529.428 & ~1.423 & ~~~8 & ~~8.74\\
& ~~~~$\ell$ = 4 & ~5893.955 & ~1625.554 & ~1.198 & ~~~5 & ~~13.06\\
\hline
\multirow{2}{*}{~~Imaginary} & ~~~~$\ell$ = 0  & ~3003.343 & ~869.994 & ~40.422 & ~~~20 & ~~0.76\\
& ~~~~$\ell$ = 2 & ~4.266 & ~0.0002 & ~0.003 & ~~~8 & ~~0.76\\
& ~~~~$\ell$ = 4 & ~99.949 & ~70.002 & ~0.033 & ~~~5 & ~~0.59\\
\hline \hline
\end{tabular}
\label{parameters}
\end{table}
The potentials corresponding to all partial waves utilising the parameters along with plots of real parts of scattering phase shifts w.r.t experimental data given in Table \ref{parameters} are given in figure \ref{real}.\\
The outer attractive and repulsive core of the potentials are comparable to the potentials given by Darriulat $et.al.$ \cite{Darriulat}. Since we are considering lab energies up-to 120 MeV, therefore we have calculated imaginary phase shifts along with real parts of phase shifts. The imaginary phase shifts along with their corresponding potentials are given in figure \ref{imag}.
The experimental data for real phase shifts have been taken from Heydenberg and Temmer, Russel, Phillips and Reich and Jones, Phillips and Miller E=0.15-9 MeV (1956-60), Nilson, Jentschke, Briggs, Kerman, Snyder E=12.3-22.9 MeV (1958), Igo E=23.1-47.1 MeV (1960),  Tombrello and Senhouse E=3.84-11.88 MeV (1963), Darriulat, Igo and Pugh E=53-120 MeV (1965) and Chen and Ronald E=18-29.5 MeV (1974) \cite{Salpha}. And for imaginary phase shifts, we have taken data from Afzal $et.al.$.\\
It is clear from the potential plot of $\ell=0$ channel for real part of phase shift given in inset of figure \ref{real}, the Coulomb barrier height comes out to be $\approx$ 0.09 MeV which corresponds to energy of quasi bound state in S-state of $\alpha-\alpha$ scattering which results due to repulsion caused by Coulomb interaction \cite{Buck}. The energy of quasi bound S-state can be observed in $\alpha-\alpha$ scattering experiments where the strong resonance of S-state occurs at 0.09184 MeV energy. 
\section{Conclusion}\label{sec4}
The real and imaginary scattering phase shifts for $\ell$ = 0 (S-channel), $\ell$ = 2 (D-channel) and $\ell$ = 4 (G-channel) have been computed up to 120 MeV by considering Malfliet-Tjon (MT) along with Hulth{\'e}n potential as screened Coulomb potential as the interaction potential. The best-fitted parameters are found to give a good match with the experimental data. Thus, PFM stands as an efficient tool for phase shift calculations in quantum mechanical scattering problems for local as well as non-local potentials. We could summarise all of our efforts into three main points: \\
(i) Malfliet-Tjon potential results in effective inverse interaction potentials for $\alpha-\alpha$ scattering for $\ell = 0,2,4$ channels.\\
(ii) MT potential can also be utilised in complex potential model
to study inelastic scattering effectively.
(iii) The Phase Function Method is an effective tool for phase shift calculations and developing suitable phenomenological potentials for a broad energy range of $\ell = 0,2,4$ partial waves.
 
\end{document}